\documentclass[a4paper,11pt]{article}
\pdfoutput=1 % if your are submitting a pdflatex (i.e. if you have
\usepackage{jcappub} % for details on the use of the package, please
\newcommand{\Dataset}{\mathcal{D}}
\newcommand{\reals}{ \mathbb{R} }
\newcommand{\trasp}{\top}
\usepackage[T1]{fontenc} % if needed
\usepackage{bm}
\usepackage{graphicx}
\usepackage{multirow}
\usepackage{subcaption}
\usepackage{hyperref}
\hypersetup{
    colorlinks=true,
    linkcolor=blue,
    filecolor=magenta,      
    urlcolor=cyan}

\newcommand{\orcid}[1]{\href{https://orcid.org/#1}{\textcolor[HTML]{A6CE39}{\aiOrcid}}}

\title{\boldmath Constraining the Reionization History  using Bayesian Normalizing Flows}

\author{H\'ector J. Hort\'ua\href{https://orcid.org/0000-0002-3396-2404}{\includegraphics[scale=0.5]{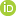}}, Luigi  Malag\`o,  Riccardo Volpi\href{https://orcid.org/0000-0003-4485-9573}{\includegraphics[scale=0.5]{ORCID-iD_icon-16x16.png}}}
\affiliation{Machine Learning and Optimization Group,\\
Romanian Institute of Science and Technology (RIST),\\
Cluj-Napoca, Romania}

\emailAdd{hortua.orjuela@rist.ro}
\emailAdd{malago@rist.ro}
\emailAdd{volpi@rist.ro}

\abstract{
The next generation 21 cm surveys open a new window onto  the early stages of cosmic structure formation and  provide  new insights about  the Epoch of Reionization (EoR).  However, the non-Gaussian nature of the 21 cm signal along with the huge amount of data generated from these surveys will require  more advanced techniques  capable to efficiently extract the necessary information to constrain the Reionization History of the  Universe.  In this paper we  present the use of Bayesian Neural Networks (BNNs) to  predict the posterior distribution for four astrophysical and cosmological parameters. Besides achieving state-of-the-art prediction performances, the proposed methods provide accurate estimation of parameters uncertainties and infer correlations among them.
Additionally, we demonstrate the advantages of Normalizing Flows (NF) combined with BNNs, being able to model more complex output distributions and thus capture key information as non-Gaussianities in the parameter  conditional density distribution for astrophysical and cosmological datasets.
Finally, we propose novel calibration methods employing Normalizing Flows after training, to produce reliable predictions, and we demonstrate the advantages of this approach both in terms of computational cost and prediction performances.}

\begin{document}
\maketitle
\flushbottom

\section{Introduction}
In the past decade, Cosmology has entered into a new precision era  due to the considerable number of experiments performed to obtain information both from early stages of the Universe through the Cosmic Microwave Background (CMB) and late times via deep redshift surveys of large-scale structures. These measurements have  yielded precise  estimates  for  the   parameters  in the standard  cosmological model, establishing  the current understanding of the Universe. However, the intermediate  time known as  Epoch of Reionization (EoR),  when the first stars and galaxies  ionized  the InterGalactic Medium (IGM), remains vastly unexplored. This period is relevant  to understand the properties of the first structures of our Universe and provide complementary information  related to fundamental Cosmology, inflationary models, and neutrino constraints, among others, e.g., \cite{Pritchard_2012}. It is known that the EoR can be studied indirectly through its imprint in the IGM, using the redshifted 21cm line~\cite{10.1093/mnras/stv571}.
This line results from the hyperfine splitting of the ground state of the hydrogen atom due to the coupled magnetic moments between the proton and the electron, emitting radiation with a 21cm wavelength, then redshifted by the expansion of the Universe~\cite{Pritchard_2012}.
Future experiments such as Hydrogen Epoch of Reionization Array (HERA)\footnote{https://reionization.org/} and the  Square Kilometre Array (SKA)\footnote{https://www.skatelescope.org/} are intended to measure this 21cm signal over a wide range of redshifts providing 3D maps of the first hundreds millions years of the Universe. These instruments are expected to generate a huge amount of spectra, encouraging the development of automated methods capable of reliably estimating physical parameters with great accuracy.
Recently, Deep Neural Networks (DNNs) have been applied in several fields of Astronomy because of their ability to extract complex information from data, which makes  it benefit  for analysing non-Gaussian signatures.  In particular, the application of DNNs to the 21 cm signal has received considerable attention due to the success of classifying reionization models~\cite{10.1093/mnras/sty3282} or estimating physical parameters~\cite{10.1093/mnras/stz010}. 
For example, in~\cite{10.1093/mnras/stz010} 2D images corresponding to  slices along the line-of-sight axis of the light-cones were used for training Convolutional
Neural Networks (CNN) in order to estimate some astrophysical parameters. More recently~\cite{La_Plante_2019} and~\cite{hassan2019constraining} generalized the previous findings by incorporating  contamination from simulated SKA-like noise.
However, DNNs are prone to  overfitting due to the high number of parameters to be adjusted, and
do not provide a measure of the uncertainty for the estimated parameters, see for instance~\cite{NIPS2011_4329,KWON2020106816,Cobb_2019}.
These limitations can be addressed by following a Bayesian approach, both intrinsically providing an effective regularization during training and allowing to quantify the uncertainty in the predicted parameters at inference time~\cite{hort2020parameters}.

In this paper, we generalize  these preliminary works  related to the application of DNNs on the 21 cm data by implementing  Bayesian Neural Network (BNNs),  in order to obtain the posterior probability estimates of the physical parameters and their correlations.
We discuss methods for calibrating uncertainties in Bayesian Networks and we propose approaches toward the efficient extraction of information in non-Gaussian signals via diffeomorphic transformations of the output distribution.

This work  is organized as follows. In Sec.~\ref{sec:VarInf} we introduce the variational inference formalism which allows to compute the aleatoric and epistemic uncertainty of BNNs, and we also comment on the use of bijectors  for improving inference tasks. In Sec.~\ref{sec:data} we describe the Reionization model used in this work and the generation of the synthetic dataset. In Sec.~\ref{sec:arch_cal} we describe the network architecture, and in Sec.~\ref{sec:results} we show the results  related to  the  potential application of BNNs to obtain approximate posteriors over the parameter space. Moreover, we  also present the  most relevant findings  about implementing Normalizing Flows in inference task and their advantages in estimating the parameters relevant for the EoR. Finally, conclusions and future works are shown in Sec.~\ref{sec:conclus}.

\section{Variational Inference}
\label{sec:VarInf}
BNNs provide the adequate groundwork to output reliable estimations for many  machine learning tasks. Let us consider a training dataset $\Dataset=\{(\bm{x}_i, \bm{y}_i)\}_{i=1}^D$ formed by $D$ couples of images $\bm{x}_i \in \reals^M $  and their respective targets $\bm{y}_i \in \reals^N$. By setting a prior distribution $p({\bm w})$ on the model parameters ${\bm w}$, the posterior distribution can be obtained from Bayes' law as $ p({\bm w}|\mathcal{D})\sim p(\mathcal{D}|{\bm w})p({\bm w})$.
Unfortunately, the  posterior usually cannot  be obtained analytically and thus approximate methods are commonly used to perform the inference task. The Variational Inference approach approximates the exact posterior  $p({\bm w}|\mathcal{D})$ by a parametric distribution $q({\bm w}|\theta)$  depending on a set of variational parameters ${\bm \theta}$~\cite{NIPS2011_4329}. These parameters are adjusted to minimize a certain loss function, usually given by the KullBack-Leibler divergence  $\text{KL}(q({\bm w}|{\bm \theta})||p({\bm w}|\mathcal{D}))$. It has been shown that minimizing the KL divergence is equivalent to minimizing the following objective  function~\cite{NIPS2011_4329}
\begin{equation}\label{eq:4}
\mathcal{F}_{VI}(\mathcal{D},{\bm \theta}) = \text{KL}(q({\bm w}|{\bm \theta})||p({\bm w}))\\
-\sum_{(\bm{x},\bm{y})\in\Dataset}\int_\Omega q({\bm w}|{\bm \theta}) \ln p({\bm y}|{\bm x},{\bm w}) d{\bm w} \; .
\end{equation}
To infer the correlations between the parameters uncertainties~\cite{2019arXiv191108508H,KWON2020106816,Gal2015BayesianCN}, we need to predict the full covariance matrix. This requires to produce in output of the last layer of the network a mean vector $\bm{\mu}\in \reals^{N}$ and a covariance matrix $\Sigma\in\reals^{N\times N}$ representing the aleatoric uncertainty, for instance parameterized through its Cholesky decomposition $\Sigma=LL^\trasp$. These outputs determine the Negative Log-Likelihood (NLL)
%
% for
when $p({\bm y}|{\bm x},{\bm w})$ is
a Multivariate Gaussian distribution~\cite{Dorta_2018,Cobb_2019,2019arXiv191108508H} 
\begin{equation}\label{eq:16}
% \mathcal{L}
\ln p({\bm y}|{\bm x},{\bm w}) \sim \frac{1}{2}\log |\Sigma|+ \frac{1}{2}(\bm{y}-\bm{\mu})^\trasp \Sigma^{-1}(\bm{y}-\bm{\mu}) \; .
\end{equation}

Let ${\hat{{\bm \theta}}}$ be the value of ${\bm \theta}$ after training, corresponding to a minimum of $\mathcal{F}_{VI}(\mathcal{D},{\bm \theta})$. The approximate predictive distribution  $q_{\hat{{\bm \theta}}}$ of $\bm{y}^*$ for a new input $\bm{x}^*$ can be rewritten as~\cite{Gal2015BayesianCN}
\begin{equation}\label{eq:approxpost_pygivenx}
  q_{\hat{{\bm \theta}}}({\bm y}^*|{\bm x}^*)=\int_{\Omega} p({\bm y}^*|{\bm x}^*,{\bm w})q({\bm w}|{\hat{\bm\theta}})d{\bm w} \; .
\end{equation}
Moreover~\cite{Gal2015Dropout} proposed an unbiased Monte-Carlo estimator for Eq.~\ref{eq:approxpost_pygivenx}
\begin{equation}\label{eq:approxpost_pygivenx_mcest}
 q_{\hat{{\bm \theta}}}({\bm y}^*|{\bm x}^*) \approx \frac{1}{K}\sum_{k=1}^K p({\bm y}^*|{\bm x}^*,\hat{{\bm w}}_k), \quad \mbox{with } \hat{{\bm w}}_k \sim q({\bm w}|{\hat{\bm\theta}}) \;,
\end{equation}
where $K$ is the number of samples.
In Bayesian deep learning~\cite{NIPS2017_7141} two main uncertainties are of interest: the aleatoric, capturing the inherent noise in the input data, and the epistemic, capturing the uncertainty in the model, typically due to the lack of data points during training which are similar to the current observation.
To obtain both uncertainties~\cite{KWON2020106816},
we can invoke the total covariance law for a fixed ${\bm x}^*$
\begin{equation}\label{eq:6}
  \mathrm{Cov}_{q_{\hat{{\bm \theta}}}}({\bm y}^*,{\bm y}^*|{\bm x}^*) = \mathbb{E}_{q_{\hat{{\bm \theta}}}}[{\bm y}^*{\bm y}^{* \trasp}|{\bm x}^*] - \mathbb{E}_{q_{\hat{{\bm \theta}}}}[{\bm y}^*|{\bm x}^*]\mathbb{E}_{q_{\hat{{\bm \theta}}}}[{\bm y}^*|{\bm x}^*]^{\trasp},
 \end{equation}
the images are forward passed through the network $T$ times, obtaining a set of mean vectors $\bm{\mu}_t$ and a covariance matrices $\Sigma_t$. Then, an estimator for the total covariance of the trained model
% Eq.~\eqref{eq:6}
can be written as
\begin{equation}\label{eq:cov}
% \widehat{\mathrm{Cov}}(\bm{y}^*,\bm{y}^*|\bm{x}^*)\approx
\mathrm{Cov}_{q_{\hat{{\bm \theta}}}}({\bm y}^*,{\bm y}^*|{\bm x}^*)
\approx
\underbrace{\frac{1}{T}\sum_{t=1}^{T}\Sigma_t}_\text{Aleatoric}+ \underbrace{\frac{1}{T}\sum_{t=1}^{T}( {{\bm{\mu}}}_{t}-\bm{\overline{\mu}})( {\bm{\mu}}_{t}-\bm{\overline{\mu}})^\trasp}_\text{Epistemic}, 
\end{equation}
with $\bm{\overline{\mu}}= \frac{1}{T}\sum_{t=1}^{T} {\bm{\mu}}_t$. In this setting, BNNs  can be used to learn the correlations between the targets and to produce estimates of their uncertainties.

\subsection{Normalizing Flows in inference}
Transforming probability distributions has become a powerful tool in deep learning. The main idea of a Normalizing Flow is to use a diffeomorphism (a differentiable and bijective mapping) to transform the sample space of a distribution. It can be demonstrated that this is equivalent to transforming the probability distribution and thus allowing for a more expressive output of the model~\cite{2018arXiv180204908T}. For a more comprehensive introduction about flows we remind the reader to~\cite{papamakarios2019normalizing,kobyzev2019normalizing}.

\subsubsection{Basics concepts of Normalizing Flows}
\label{sec:finite_flows}

Let us consider $\bm{u}\in U\subset \reals^{D}$ and $\bm{x}\in X\subset \reals^{D}$ two $D$ dimensional vectors in some subsets of $\reals^{D}$. We can define probability distributions for  $\bm{u}$ and $\bm{x}$, in the associated sample spaces $U$ and $X$, respectively.
Let $f$ be a diffeomorphism between the two sample spaces $f: U \rightarrow X$. Knowing the probability distributions $q_u$, we can define $q_x$ as
\begin{equation}
  q_x(\bm{x}) = q_u(\bm{u})\left| \det  \frac{\partial f(\bm{u})}{\partial \bm{u}} \right|^{-1}, \label{eq:drule}
\end{equation}
where $\bm{x}=f(\bm{u})$. We can construct  more complex densities by applying successively the bijector~\eqref{eq:drule}, thus transforming an initial random variable $\bm{u}$ with distribution $q_{u}$ ($=q_{0}$) through a series of transformations $f_{n}$ as
\begin{align}
 \bm{x}_n &= f_{n} \circ \ldots \circ f_2 \circ  f_{1} (\bm{u})  \label{eq:fnComposition} \\
  \ln q_{n} (\bm{x}_n) &=
  \ln  q_{u} (\bm{u}) -\sum_{i=1}^{n} \ln \left| \det \frac{\partial f_i(\bm{x}_{i-1})}{\partial \bm{x}_{i-1} }\right|,
  \label{eq:logdetinv} 
% \label{eq:nested}
\end{align}
where we defined $\bm{x}_0 = \bm{u}$ for convenience of notation.
Any expectation  $\mathbb E_{q_{n}}[h(\bm{x}_n)]$ can be written as an expectation under $q_{u}$ as
\begin{equation}
\mathbb{E}_{q_{n}}[h(\bm{x}_n)] = \mathbb{E}_{ q_u }[ h( f_{n} \circ f_{n-1} \circ   \ldots   \circ  f_{1} ( \bm{u} ) ]\;.
\label{eq:expectations}
\end{equation}

Through a suitable choice of the diffeomorphisms $f_n$, we can start from simple factorized distributions such as a mean-field Gaussian and apply normalizing flows  to obtain complex and multi-modal distributions~\cite{huang2018neural,papamakarios2019normalizing,kobyzev2019normalizing}.

Ideally a flow is both expressive and requires a reduced additional cost. Several transformations have been proposed in the literature~\cite{papamakarios2019normalizing,kobyzev2019normalizing} to compute efficiently both \eqref{eq:fnComposition} and \eqref{eq:logdetinv}. An example are Masked Autoregressive Flows~\cite{papamakarios2017masked} (MAF) and its inverse, the Inverse Autoregressive Flows~\cite{kingma2016improved} (IAF)  which allow to compute  efficiently Eq.~\eqref{eq:logdetinv}   and Eq.~\eqref{eq:fnComposition} respectively. Real valued non-volume preserving transfromations~\cite{dinh2016density} (NVP) are a special case of both MAF and IAF, in which $d$ pass-through units are selected and the transformations on the other units are function of these pass-through units.
A more recent improvement is represented by neural ODE~\cite{chen2018neural,grathwohl2018ffjord} in which the flow is represented by a continuous transformation specified through an ordinary differential equation.
In this paper we will use: NVP, MAF and IAF as a proof of concept to show how 
to provide a more flexible and scalable distribution in the output of the network with the purpose of extracting complex  features  in the data such as a non-Gaussianities and provide well calibrated BNN model.
\section{Dataset generation}\label{sec:data}
We generated 21cm simulations through the semi-numerical code   21cmFast~\cite{10.1111/j.1365-2966.2010.17731.x}, producing realizations of halo distributions and ionization maps at high redshifts.  Through  approximate methods,  the  code    generates  full  3D  realizations  of  the  density,  ionization,  velocity,  spin  temperature, and  21-cm brightness temperature fields. The latter is computed as~\cite{FURLANETTO2006181}
\begin{equation}\label{tb21cm}
    \delta T_b\approx 27(1+\delta_m)x_{HI}\Big(\frac{T_S-T_\gamma(z)}{T_S} \Big)\Big(\frac{\Omega_b h^2}{0.023} \Big)\Big(\frac{1+z}{10}\frac{0.15}{\Omega_m h^2} \Big)^{\frac{1}{2}}\,\text{mK},
\end{equation}
where $T_S$ and $T_\gamma (z)$ are  the gas spin  and the CMB temperatures at redshift $z$ respectively, $\delta_m$ is the density contrast of baryons, and $x_{HI}$ denotes the neutral fraction of hydrogen.
At $z\approx6-20$, the scattering  of  photons from  the  first  galaxies  and  black holes warms up the IGM enhancing the ionization  fraction~\cite{10.1093/mnras/stz3107}. Once the gas is ionized at  some percent level ($\sim 25\%$, see~\cite{Santos_2008}), the spin temperature becomes  greater than CMB temperature, $T_S>>T_\gamma$, and  thus,  the  dependence  on $T_S$ may be neglected  in Eq.~\ref{tb21cm}. The contrast density which depends on cosmological parameters is found by using the same techniques employed in numerical cosmological simulations~\cite{FURLANETTO2006181,Santos_2008}, while the HI ionized field parametrized by $x_{HI}$ and dependent on different astrophysical parameters, is obtained from the excursion-set approach~\cite{10.1111/j.1365-2966.2010.17731.x}. At the end, the combination of the numerical solutions for the contrast density and $x_{HI}$, allows to compute the  21 cm brightness temperature field via Eq.~\ref{tb21cm}.
In order to generate the synthetic dataset for training the network, we follow the ideas started in~\cite{hort2020parameters} where  we have varied four parameters. Two parameters corresponding to the cosmological context: the matter density parameter $\Omega_m \in [0.2,0.4]$, and the amplitude of mass fluctuations on $8h^{-1}$Mpc, $\sigma_8 \in [0.6,0.8]$. The other two parameters corresponding to the astrophysical context: the ionizing efficiency of high-z galaxies $\zeta \in [10,100]$ and the minimum virial temperature of star-forming haloes $T^F_{vir} \in [3.98,39.80]\times10^4$K (hereafter represented in log10 units). For each set of parameters we produce 20 images at different redshifts in the range $z \in [6,12]$, and we stack these redshift-images into a single multi-channel tensor. This scheme brings two main advantages, first the network can extract effectively the information encoded over images as it was reported in \cite{2019arXiv191108508H}, and secondly, it represents adequately the signals for the  next-generation interferometers and provides advantages when we need to include effects of foreground contamination~\cite{La_Plante_2019}.
As a final result we have obtained $6,000$ images each with size of $(128,128,20)$ and resolution $1.5$ Mpc. We used a 70-10-20 split for training, validation and test, respectively.

\section{Architecture and network calibration}\label{sec:arch_cal}
All the networks are implemented using  TensorFlow~\footnote{\MakeLowercase{h}ttps://www.tensorflow.org/} and TensorFlow-Probability~\footnote{\MakeLowercase{h}ttps://www.tensorflow.org/probability}. We used a modified version of the VGG architecture with 5 VGG blocks (each made by two Conv2D layers and one max pooling) and channels size [32, 32, 32, 32, 64]. Kernel size is fixed to 3$\times$3 and activation function used is LeakyReLU ($\alpha=-0.3$). Each convolutional layer in the network is followed by a batch renormalization layer. The last layer is dense with output corresponding to the mean of predictions $\mu$ and a lower triangular matrix ${L}$, cf.~\cite{Dorta_2018,Cobb_2019,2019arXiv191108508H}, yielding a multivariate Gaussian distribution with mean $\bm{\mu}$ and covariance $\Sigma={L}{L}^\top$ to guarantee positive definiteness. To model the distribution over the weights we used two methods, Dropout~\cite{Gal2015Dropout} and Flipout~\cite{wen2018flipout}. For the Dropout method, after each Conv2D  we included a dropout layer which is always turned on, during both training and test. In ~\cite{Gal2015Dropout} the authors found that this method is equivalent to work with BNNs i.e., the weights are sampled from a (kind of) Bernoulli distribution and during inference allows to obtain uncertainties of the output's network. On the other hand, we used the Flipout method, assuming in Eq.~\ref{eq:4} a Gaussian distribution over the weights for both prior and posterior, and providing an efficient way to draw pseudo-independent weights for different elements in a single batch~\cite{wen2018flipout}.
We trained the networks for 180 epochs with batches of 32 samples, using 10 samples from the approximate posterior for the estimation of Eq.~\ref{eq:approxpost_pygivenx_mcest} (experiments with 1 sample are also reported in Appendix for comparison). After training, to obtain the prediction distributions and the related uncertainties, we feed each input image from the test set 2,500 times to each network.
\subsection{Calibration in BNNs}\label{calibrat}
Predicting  reliable  uncertainties is crucial for classification and regression  models in many applications. However, it is known that  DNNs trained with NLL may produce poor uncertainty estimates~\cite{2017arXiv170604599G}. Weight decay,  Batch Normalization and  the choice of specific divergences in VI have been shown to be important factors influencing the calibration~\cite{2017arXiv170604599G,2017arXiv170302914L}. One way to observe this miscalibration is computing the coverage probability in the test set. For doing this,  we have binned the samples drawn from inference and computed their  mode~\cite{Perreault_Levasseur_2017}. With this value, and  assuming an unimodal posterior, we estimated the intervals that include the 68, 95, and 99$\%$ of the samples.

To solve the miscalibration of the network, we can proceed in two different ways. The first option is to calibrate the network during training by tuning hyper-parameters such as dropout rate in Dropout~\cite{2019arXiv191108508H,Perreault_Levasseur_2017}, the regularization for the scale of the variational distribution in Flipout~\cite{2019arXiv191108508H}, or  the $\alpha$-value in  the alpha-divergence energy~\cite{2017arXiv170302914L}. The second option involves post-processing techniques applied after training. For the latter option, we retrain  the last layer of the network as it was proposed in~\cite{2019arXiv191108508H}, minimizing the NLL Eq.~\ref{eq:16} distorted by  Normalizing Flows. Other methods have been reported in literature for calibration~\cite{2017arXiv170604599G,levi2019evaluating,2019arXiv191108508H}. However, for practical reasons only a selection of calibration methods which from preliminary experiments we found working best will be reported.

\section{Results}\label{sec:results}
We quantify the performance of the networks by the coefficient of determination and the accuracy of the uncertainties (well calibrated networks).
The coefficient of determination is defined as
\begin{equation}
R^2=1-\frac{\sum_i (\bm{\bar{\mu}}(\bm{x}_i)-\bm{y}_i)^2}{\sum_i (\bm{y}_i-\bm{\bar{y}})^2}
\end{equation}
where $\bm{\bar{\mu}}(\bm{x}_i)$ (see Eq.~\ref{eq:cov}) is the prediction of the trained Bayesian network , $\bm{\bar{y}}$ is the average of the true parameters and the summations are performed over the entire test set. $R^2$ ranges from 0 to 1, where 1 represents perfect inference. Regarding uncertainties accuracy, for calibrated networks $\bm{y}_i$ should fall in a $\beta\%$ confidence interval of the conditional density estimation~\eqref{eq:approxpost_pygivenx_mcest} approximately $\beta\%$ of the time, where $\beta=\{68.3, 95.5, 99.7\}$ corresponding to 1, 2, and 3$\sigma$ confidence levels of a normal distribution.
\subsection{Comparison among BNNs methods}\label{sec:bnncomparison}
For Dropout we tested several dropout rates in the range [$0.01$, $0.1$] keeping L2 regularization fixed to $1e^{-5}$, while for Flipout we tested several L2 regularizations in the range [$1e^{-5}$, $1e^{-7}$]. A detailed summary of the experiments is reported in Fig.~\ref{fig:all4CI} and Tables~\ref{table:Dropoutall}-\ref{table:Flipoutall} in Appendix.  
\begin{table}[h]
\begin{center}
\begin{tabular}{|c|c|c|c|c|c|c|c|c|}
\hline
% \multicolumn{9}{|c|}{Metrics for the best experiments}\\ \hline
  %& \multicolumn{4}{c|}{Flipout} 
  %& \multicolumn{4}{c|}{Dropout} \\ %\cline{2-9}
    &  \multicolumn{4}{c|}{Flipout (NLL=-2.94)}    & \multicolumn{4}{c|}{Dropout (NLL=-0.74)} \\\cline{2-9}
 & $\sigma_8$    &  $\Omega_m$   &  $\zeta$    & $T^F_{vir}$    &   $\sigma_8$   &   $\Omega_m$   & $\zeta$    & $T^F_{vir}$    \\ \hline
$R^2$                & 0.94 &0.98     &0.87      &0.97     &    0.87  &0.94      &0.65     &0.92    \\ \hline
C.L. $68.3\%$                &   69.6 &73.6  &72.8     &76.1      & 70.4     &  67.3    &58.5  &76.1  \\ \hline
C.L. $95.5\%$                &   96.0   &97.1      &97.4      &96.7      &  95.7    &    96.3  & 91.7     & 98.5     \\ \hline
C.L. $99.7\%$                &  99.6    &99.9      &99.7      &99.7      & 99.6     &    99.8  & 99.8     &99.9      \\ \hline
\end{tabular}
\caption{Metrics for the best experiments with Flipout and Dropout.}
\label{table:bestCI}
\end{center}
\end{table}
\begin{figure}[htb!]
\begin{center}
%\framebox[4.0in]{$\;$}
\includegraphics[width=0.7\linewidth]{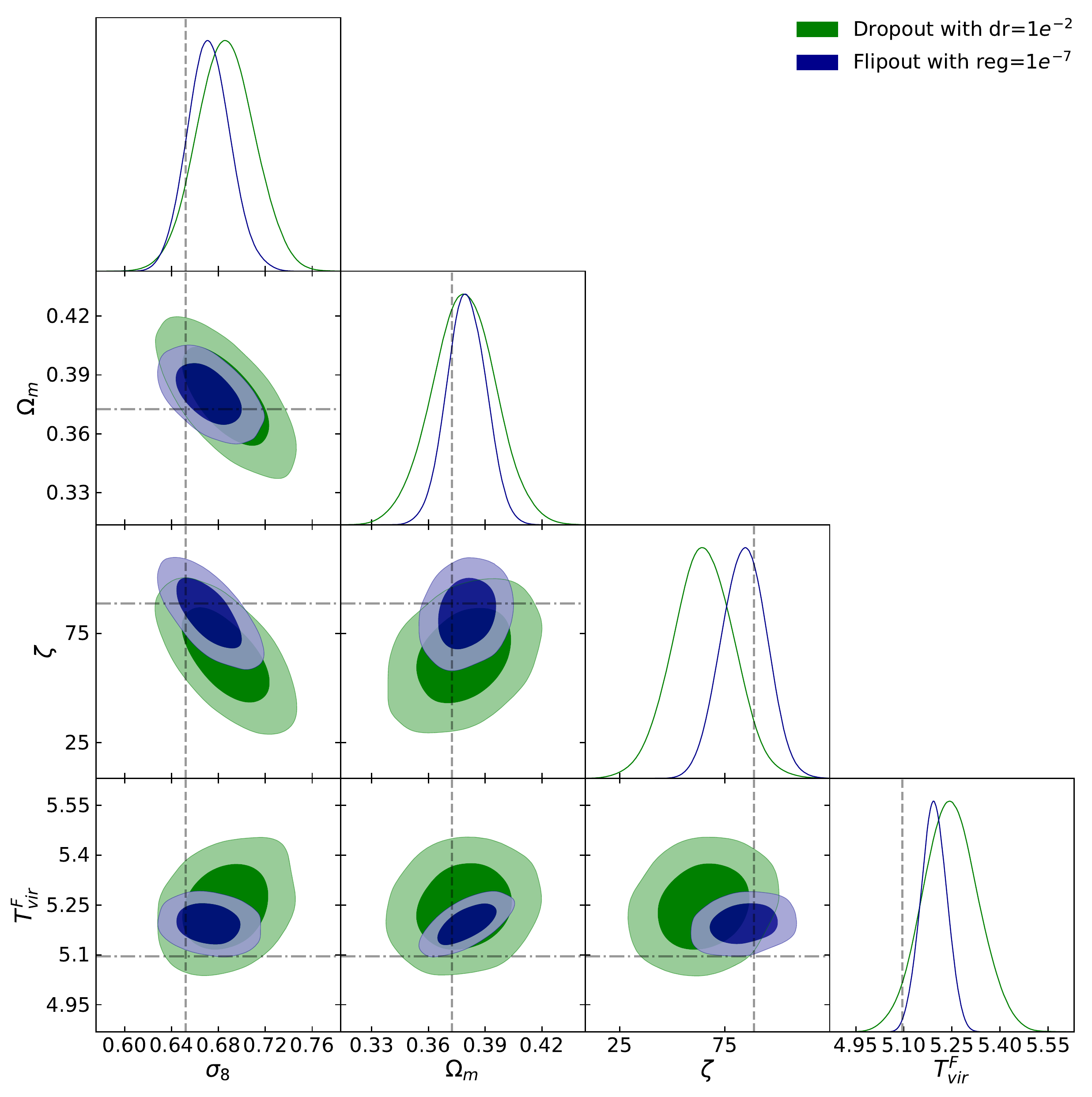}
%\fbox{\rule[-.5cm]{0cm}{4cm}\rule[-.5cm]{4cm}{0cm}}
\end{center}
\caption{Posterior distributions of the parameters for one example in the test set. The dashed lines stand for the real values. The contour regions in the two-dimensional posteriors stand for 68 and 95$\%$ confidence levels.}
\label{fig:bestCIsp}
\end{figure}
Table~\ref{table:bestCI} reports the best configuration of the network: Flipout with L2 regularizer $1e^{-7}$ and Dropout with dropout rate $1e^{-2}$. We report in the same table the coefficient of determination and average confidence intervals for Flipout and Dropout after calibration~\cite{2019arXiv191108508H}. Flipout obtains the best estimations even though tends to overestimate its uncertainties. The results obtained are comparable with the state of the art~\cite{10.1093/mnras/stz010,hassan2019constraining} even if the architecture used in the present paper is considerably smaller, furthermore as discussed in this section our approach has the advantage to be able to provide  uncertainty estimation for the predicted parameters.

\begin{table}[h]
\centering
\begin{tabular}{|l|c|c|c|c|}
% \hline
% \multirow{}{}{} & \multicolumn{4}{l|}{Flip} \\
\hline 
                  & $\sigma_8$     & $\Omega_m$    &  $\zeta$   & $T^F_{vir}$    \\ \hline
Flipout              &   $0.672^{+0.037}_{-0.036}   $   &  $0.380^{+0.020}_{-0.020}   $    &   $84.00^{+20.00}_{-20.00}            $   & $5.193^{+0.080}_{-0.079}   $     \\ \hline
Dropout              &  $0.686^{+0.048}_{-0.048}$    &   $0.379^{+0.033}_{-0.033}$   &  $65.00^{+30.00}_{-30.00}$    & $5.250^{+0.170}_{-0.170}$      \\ \hline
Example true value  & $0.652$ &$0.372$ &$88.847 $&$5.096$\\ \hline
\end{tabular}
\caption{Limits at the $95\%$ confidence level of the credible interval of predicted parameters.}
\label{table:bestCIspexample}
\end{table}

In Fig.~\ref{fig:bestCIsp} we report the confidence intervals\footnote{We use the getdist~\cite{Lewis:2019xzd} package to produce the plots.} for a single example in the test set and in Table~\ref{table:bestCIspexample} we present the parameters predictions at the $95\%$ confidence level. Notice that Flipout  yields more accurate inferences and provides tighter constraints contours, see for example $T^F_{vir}$-$\zeta$. Moreover, the correlations extracted from EoR such as $\sigma_8$-$\Omega_m$ (see Fig.~\ref{fig:bestCIsp}) provide significant information for breaking  parameter degeneracies   and thus,  be able  to improve the existing  measurements on cosmological parameters~\cite{Pritchard_2012,McQuinn_2006,ZHANG2019135064}. In the following sections we will focus on  the methods used  for producing reliable uncertainties.
Since  Flipout  achieves better performances than Dropout, from now on we will use Flipout to determine the performance of the  subsequent calibration experiments.

\subsection{Normalizing Flows during Training}
A good predictive distribution depends on how  well the parametric distribution  matches  the  exact  posterior.
For the case presented above, the variational distribution provides a simple Gaussian approximation for the conditional density $p(\bm{y}|\bm{x},\bm{w})$.

\begin{table}[h!]
\begin{center}
\begin{tabular}{|c|c|c|c|c||c|c|c|c||c|c|c|c|}
\hline
% \multicolumn{9}{|c|}{Metrics for the best experiments}\\ \hline
  %& \multicolumn{4}{c|}{Flipout} 
  %& \multicolumn{4}{c|}{Dropout} \\ %\cline{2-9}
    &  \multicolumn{4}{c|}{IAF (NLL=-3.63)}    & \multicolumn{4}{c|}{MAF (NLL=-3.19)}& \multicolumn{4}{c|}{NVP (NLL=-2.00)} \\\cline{2-13}
 & $\sigma_8$    &  $\Omega_m$   &  $\zeta$    & $T^F_{vir}$    &   $\sigma_8$   &   $\Omega_m$   & $\zeta$    & $T^F_{vir}$    &   $\sigma_8$   &   $\Omega_m$   & $\zeta$    & $T^F_{vir}$ \\ \hline
$R^2$                & 0.93 &0.97     &0.86      &0.97     &    0.93  &0.98      &0.86     &0.97    &0.93&0.96&0.86&0.97\\ \hline
C.L. $68.3\%$                &   65.6 &69.8  &64.0     &66.8      & 65.6     &  66.0    &60.3  &67.7  &56.6&71.1&67.1&68.9\\ \hline
C.L. $95.5\%$                &   94.0   &95.2      &93.0      &94.0      &  95.2    &    94.0  & 90.0     & 94.3    &86.1&96.4&94.6&95.2 \\ \hline
C.L. $99.7\%$                &  99.1    &99.7      &98.7      &99.3      & 99.4     &    99.2  & 97.8     &99.3     &96.4&99.6&99.5&99.4 \\ \hline
\end{tabular}
\caption{Metrics for the best experiments with Normalizing Flows.}
\label{table:bestCIflow}
\end{center}
\end{table}
Normalizing Flows map an initial probability distribution through a series of transformations to produce a richer, and even a multi-modal distribution~\cite{2018arXiv180204908T}.

We consider different kinds of Normalizing Flows acting on the output distribution of a BNN: the inverse autoregressive flow (IAF), Masked Autoregressive Flow (MAF) and non-volume preserving flows (NVP). The results of these experiments are reported in Table~\ref{table:bestCIflow}. We observe that the $R^2$ are comparable for all methods, but the NLL is higher for the IAF, which means that this method tends to recover better accuracy in the uncertainties. This is consistent  with the findings showed in the last three rows in Table~\ref{table:bestCIflow}, where the coverage probability in the test set is closer to the confidence intervals for IAF.
\begin{figure}[h!]
\begin{center}
%\framebox[4.0in]{$\;$}
\includegraphics[width=0.7\linewidth]{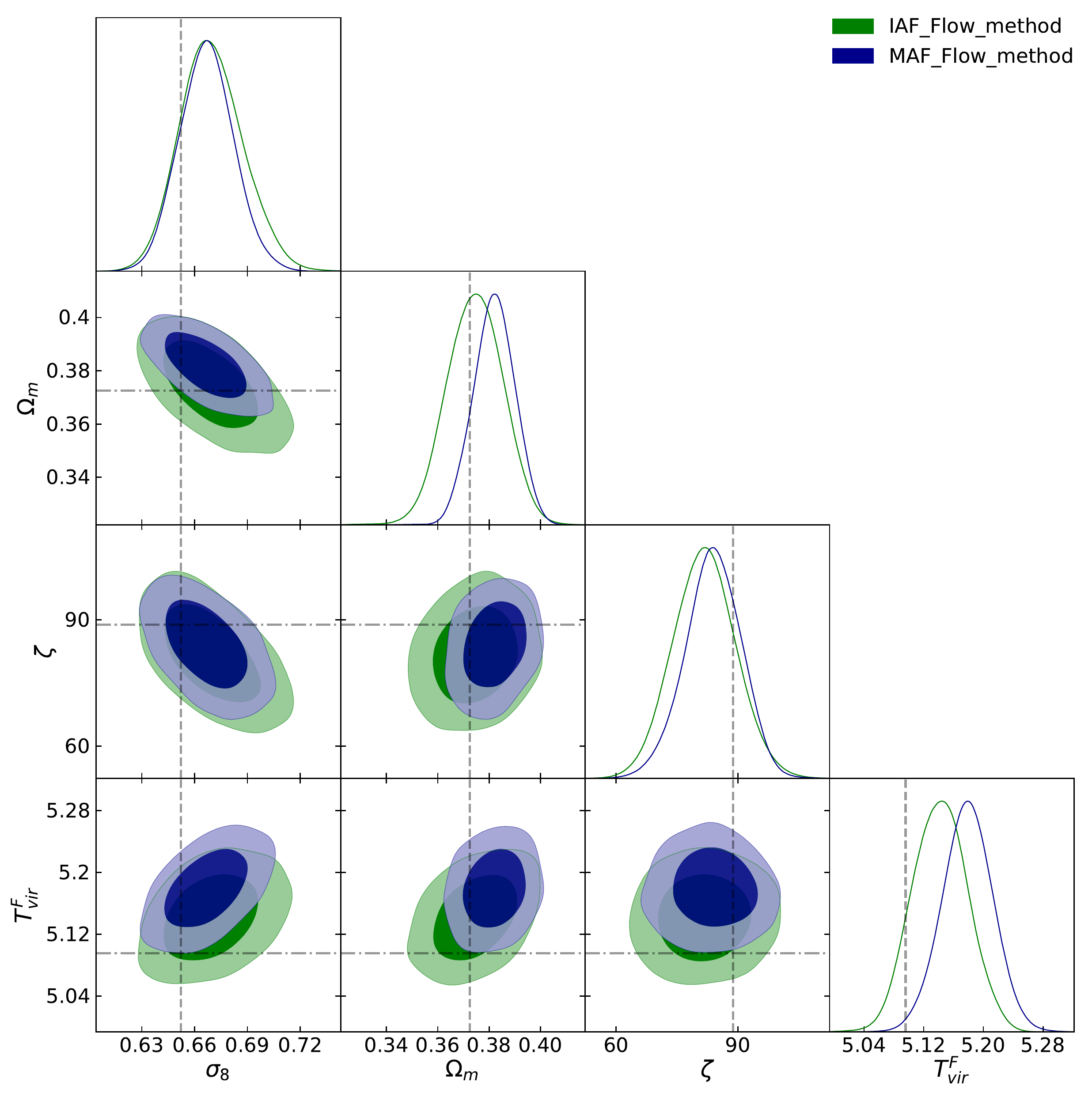}
%\fbox{\rule[-.5cm]{0cm}{4cm}\rule[-.5cm]{4cm}{0cm}}
\end{center}
\caption{Posterior distributions of the parameters for one example in the test set using Normalizing flows. The dashed lines stand for the real values. The contour regions in the two-dimensional posteriors stand for 68 and 95$\%$ confidence levels.}
\label{fig:bestCIspflow}
\end{figure}
\begin{table}[h]
\centering
\begin{tabular}{|l|c|c|c|c|}
% \hline
% \multirow{}{}{} & \multicolumn{4}{l|}{Flip} \\
\hline 
                  & $\sigma_8$     & $\Omega_m$    &  $\zeta$   & $T^F_{vir}$    \\ \hline
IAF              &   $0.670^{+0.037}_{-0.033}   $   &  $0.375^{+0.021}_{-0.021}   $    &   $82.00^{+20.00}_{-10.00}            $   & $5.142^{+0.072}_{-0.070}   $     \\ \hline
MAF              &  $0.667^{+0.031}_{-0.029}$    &   $0.382^{+0.015}_{-0.016}$   &  $84.00^{+10.00}_{-10.00}$    & $5.179^{+0.066}_{-0.068}$      \\ \hline
Example true value  & $0.652$ &$0.372$ &$88.847 $&$5.096$\\ \hline
\end{tabular}
\caption{Limits at the $95\%$ confidence level of the credible interval of predicted parameters using Normalizing Flows.}
\label{table:bestexamspflows}
\end{table}

Normalizing Flows lead to more expressive output distributions that focus on obtaining calibrated probabilities rather than enhance the precision in the target value (quantified by $R^2$) as we compare  with the ones reported in Table~\ref{table:bestCI}. In Fig.~\ref{fig:bestCIspflow} we compare the best models i.e., IAF and MAF. We can notice how smaller are the confidence regions predicted by MAF with respect to IAF, although both methods predict roughly the same orientations as expected. Finally, the values of the parameters with confidence levels at $95\%$ are reported in Table~\ref{table:bestexamspflows}.
Comparing these results with Table~\ref{table:bestCIspexample}
we can notice how the predicted parameters uncertainties are tighter and also more asymmetric, which is a consequence of the increased flexibility of the output posterior.

\subsection{Normalizing Flows in the post-process calibration}
In this part we will focus on post-process methods for network calibration using Flows.
We apply Normalizing Flows on the output distribution of the Flipout experiment reported in Sec.~\ref{sec:bnncomparison}, trained with a vanilla Multivariate Gaussian in output. We retrain the last layer as it was suggested in~\cite{2019arXiv191108508H}, plus we train the parameters of the flow. This method has the advantage that does not require  to retrain the entire network, thus demonstrating to be very cost efficient.
\begin{table}[h!]
\begin{center}
\begin{tabular}{|c|c|c|c|c||c|c|c|c||c|c|c|c|}
\hline
% \multicolumn{9}{|c|}{Metrics for the best experiments}\\ \hline
  %& \multicolumn{4}{c|}{Flipout} 
  %& \multicolumn{4}{c|}{Dropout} \\ %\cline{2-9}
    &  \multicolumn{4}{c|}{IAF (NLL=-3.80)}    & \multicolumn{4}{c|}{MAF (NLL=-3.73)}& \multicolumn{4}{c|}{NVP (NLL=-3.44)} \\\cline{2-13}
 & $\sigma_8$    &  $\Omega_m$   &  $\zeta$    & $T^F_{vir}$    &   $\sigma_8$   &   $\Omega_m$   & $\zeta$    & $T^F_{vir}$    &   $\sigma_8$   &   $\Omega_m$   & $\zeta$    & $T^F_{vir}$ \\ \hline
$R^2$                & 0.94 &0.98     &0.87      &0.98     &    0.94  &0.98      &0.87     &0.98    &0.94&0.98&0.87&0.98\\ \hline
C.L. $68.3\%$                &   66.0 &64.0  &69.2     &65.4      & 64.7     &  63.7    &69.1  &65.0  &65.9&64.8&68.8&66.0\\ \hline
C.L. $95.5\%$                &   94.0   &94.0      &95.0      &94.0      &  93.3    &    94.2  & 95.1     & 94.0    &93.0&94.0&94.0&93.0 \\ \hline
C.L. $99.7\%$                &  99.2    &99.2      &99.5      &99.6      & 99.0     &    99.3  & 99.3     &99.4     &99.0&99.2&99.0&99.0 \\  \hline
\end{tabular}
\caption{Metrics for the best experiments with Normalizing Flows after calibration.}
\label{table:bestflowcal}
\end{center}
\end{table}
\begin{figure}[h!]
\begin{center}
%\framebox[4.0in]{$\;$}
\includegraphics[width=0.7\linewidth]{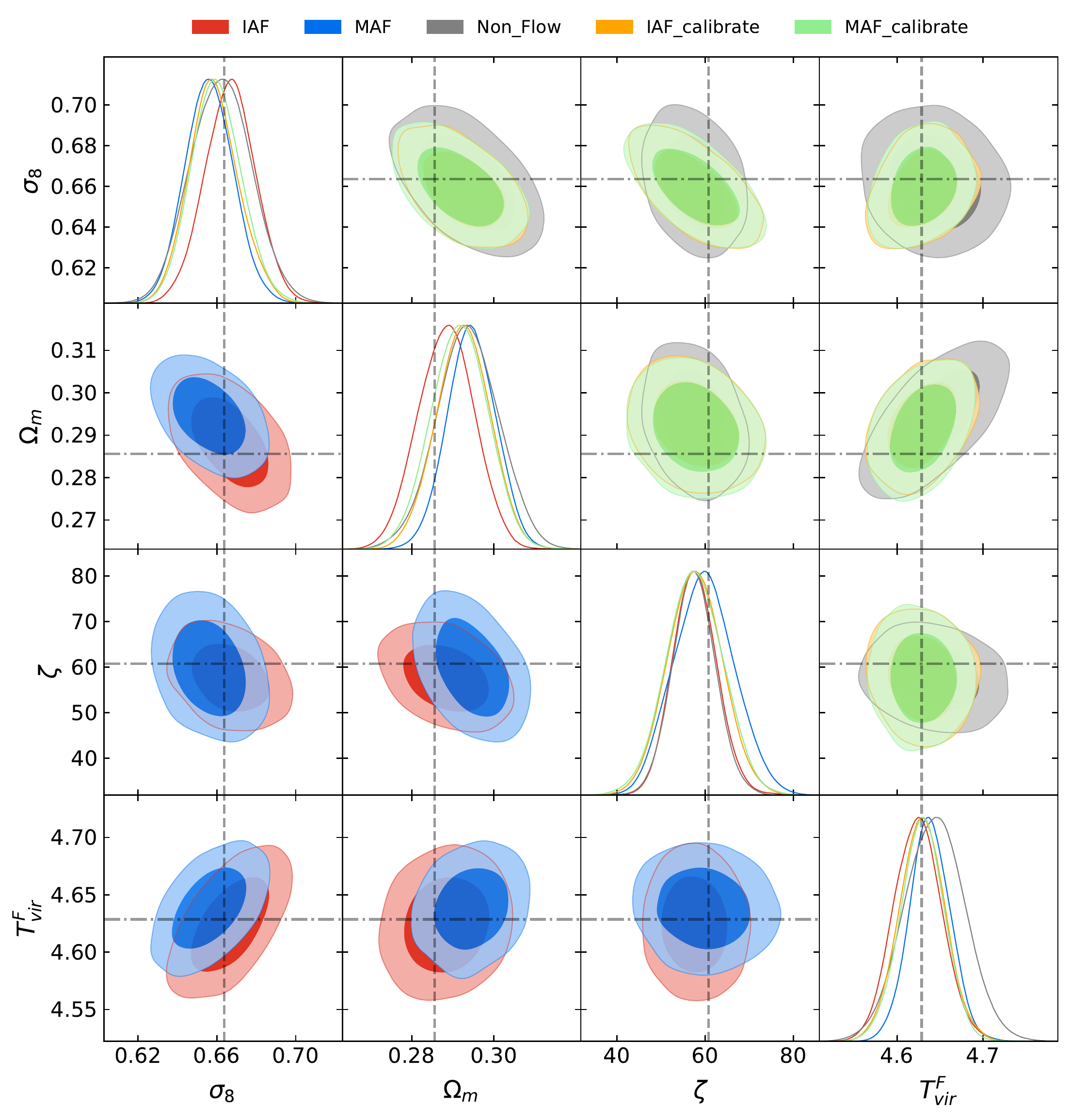}
%\fbox{\rule[-.5cm]{0cm}{4cm}\rule[-.5cm]{4cm}{0cm}}
\end{center}
\caption{Posterior distributions of the parameters for one example in the test set using Normalizing flows after calibration. The dashed lines stand for the real values. The contour regions in the two-dimensional posteriors stand for 68 and 95$\%$ confidence levels.}
\label{fig:bestCIspflowcal}
\end{figure}

Results after 120 epochs of the proposed recalibration are reported in Table~\ref{table:bestflowcal}.
Notably, the post-processing calibration outperforms all other experiments done so far, in terms of both the expected deviation from the target value, $R^2$,  and the NLL. Additionally, the improvement of the NLL leads to better calibrated networks, validated thought the coverage probabilities in Table~\ref{table:bestflowcal}. What is perhaps even more interesting is that using any Normalizing Flow method in the post-process period, outperforms the performances obtained by that same flow during training, which leads to a powerful method for obtaining models with correct interpretation of its uncertainty estimates. Calibrating with a NF results to be an easier optimization, converging to better results.

\begin{table}[h]
\centering
\begin{tabular}{|l|c|c|c|c|}
% \hline
% \multirow{}{}{} & \multicolumn{4}{l|}{Flip} \\
\hline 
                  & $\sigma_8$     & $\Omega_m$    &  $\zeta$   & $T^F_{vir}$    \\ \hline
IAF              & $0.667^{+0.025}_{-0.026}   $   &  $0.288^{+0.014}_{-0.013}    $    & $58.00^{+10.00}_{-10.00}               $   & $4.624^{+0.054}_{-0.053}   $     \\ \hline
MAF              & $0.656^{+0.025}_{-0.024}$    &   $0.295^{+0.012}_{-0.011}$   &  $60.00^{+10.00}_{-10.00}$    &  $4.638^{+0.046}_{-0.047}$      \\ \hline
NVP              &   $0.656^{+0.032}_{-0.030}   $   &   $0.302^{+0.013}_{-0.013}   $    &   $55.00^{+10.00}_{-10.00}            $   &$4.655^{+0.058}_{-0.065}  $     \\ \hline
IAF calibrated               &   $0.659^{+0.026}_{-0.024}   $   &  $0.293^{+0.012}_{-0.013}   $    & $58.00^{+10.00}_{-10.00}              $   & $4.629^{+0.053}_{-0.052}   $     \\ \hline
MAF calibrated              & $0.660^{+0.025}_{-0.024}   $   &  $0.292^{+0.013}_{-0.014}    $    &   $58.00^{+10.00}_{-10.00}            $   &$4.629^{+0.049}_{-0.051}    $     \\ \hline
NVP calibrated              & $0.663^{+0.026}_{-0.024} $   &    $0.291^{+0.013}_{-0.013}$   &  $57.00^{+10.00}_{-10.00}$    & $4.634^{+0.045}_{-0.045} $      \\ \hline
Non-Flow              &  $0.662^{+0.031}_{-0.029}    $   &  $0.294^{+0.015}_{-0.015}   $    &   $57.00^{+12.00}_{-12.00}            $   & $4.644^{+0.069}_{-0.068}  $     \\ \hline
Example true value  & $0.664$ &$0.285$ &$60.750 $&$4.629$\\ \hline
\end{tabular}
\caption{Limits at the $95\%$ confidence level of the credible interval of predicted parameters using Normalizing Flows with and without calibration.}
\label{table:bestcalflowcal}
\end{table}

\begin{figure}[h!]
    \centering
    \begin{subfigure}[b]{0.45\textwidth}
        \includegraphics[width=\textwidth]{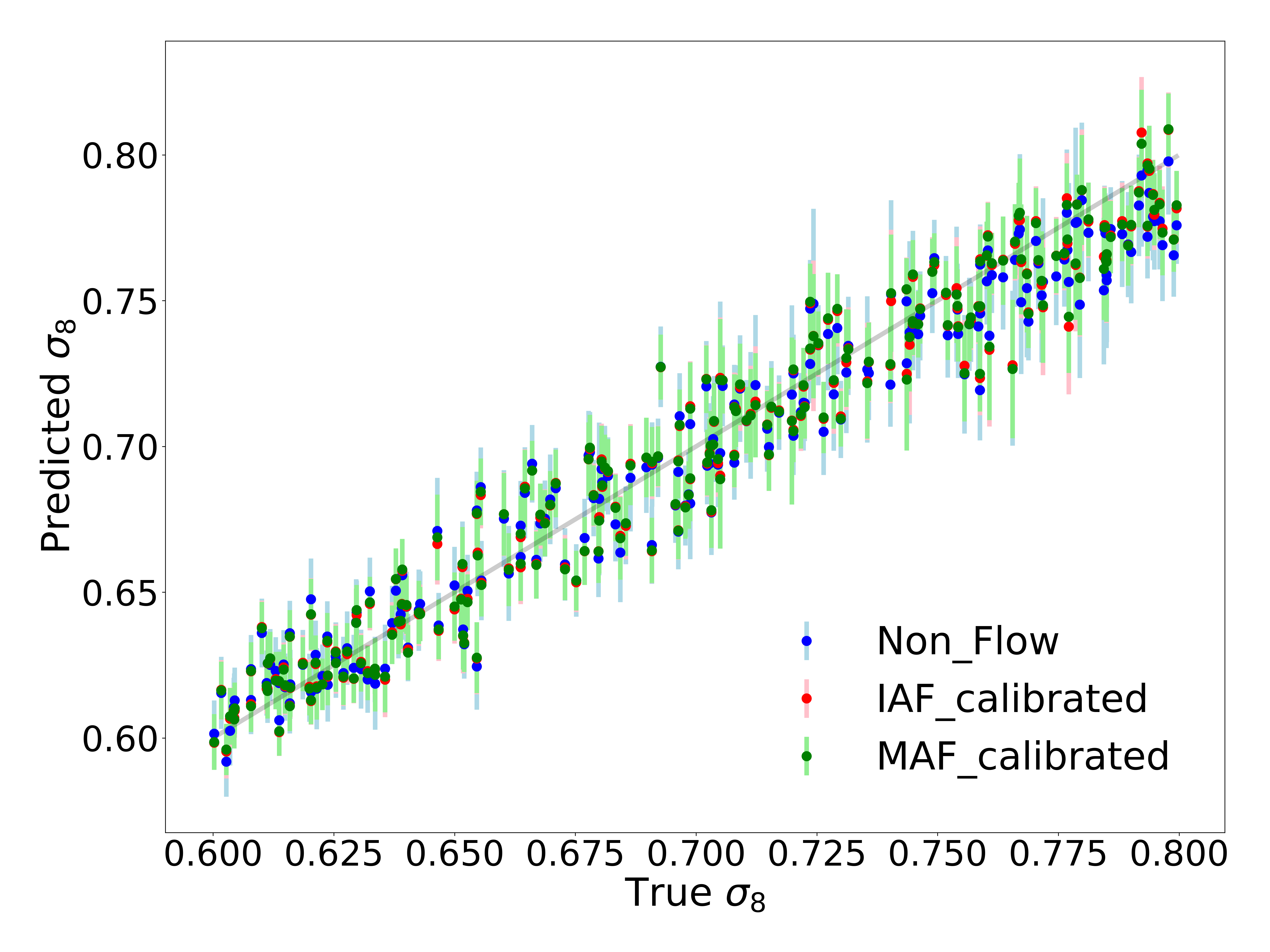}
        \caption{Predicted $\sigma_8$ value vs true value.}
        \label{figpreda}
    \end{subfigure}
    ~ %add desired spacing between images, e. g. ~, \quad, \qquad, \hfill etc. 
    \begin{subfigure}[b]{0.45\textwidth}
        \includegraphics[width=\textwidth]{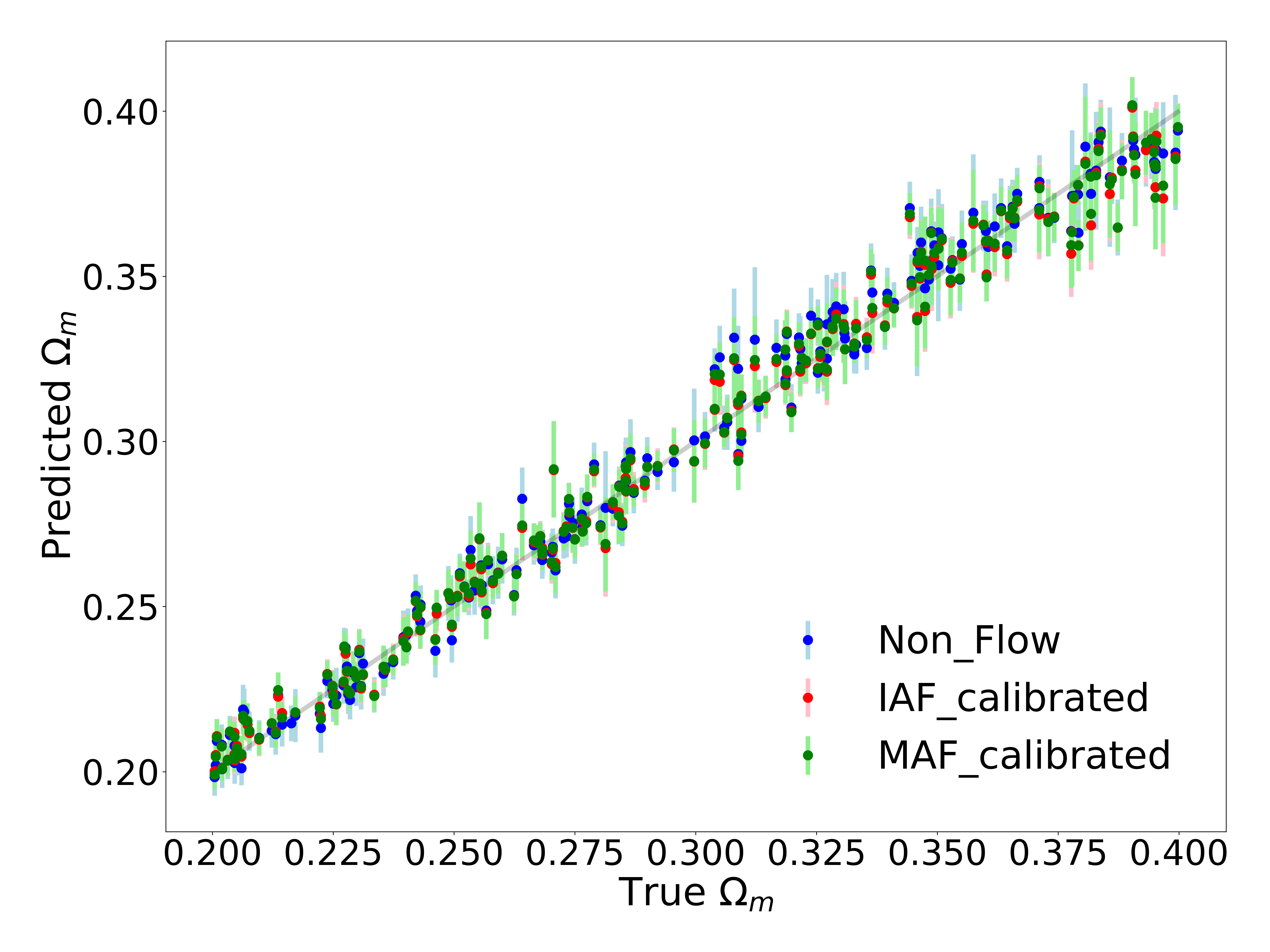}
        \caption{Predicted $\Omega_m$ value vs true value.}
        \label{figpredb}
    \end{subfigure}
    ~ %add desired spacing between images, e. g. ~, \quad, \qquad, \hfill etc. 
    \begin{subfigure}[b]{0.45\textwidth}
        \includegraphics[width=\textwidth]{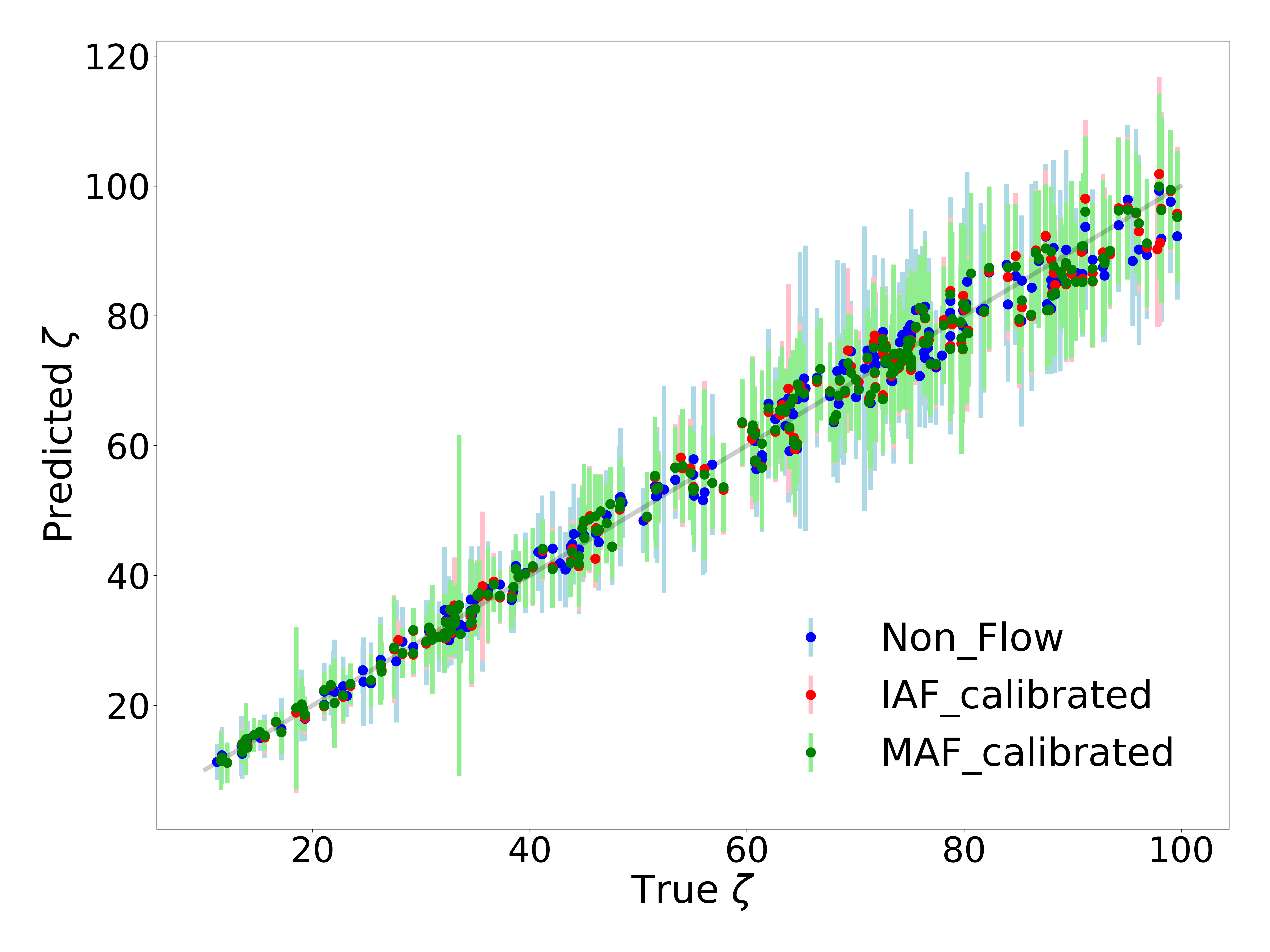}
        \caption{Predicted $\zeta$ value vs true value.}
        \label{plots}
    \end{subfigure}
     \begin{subfigure}[b]{0.45\textwidth}
        \includegraphics[width=\textwidth]{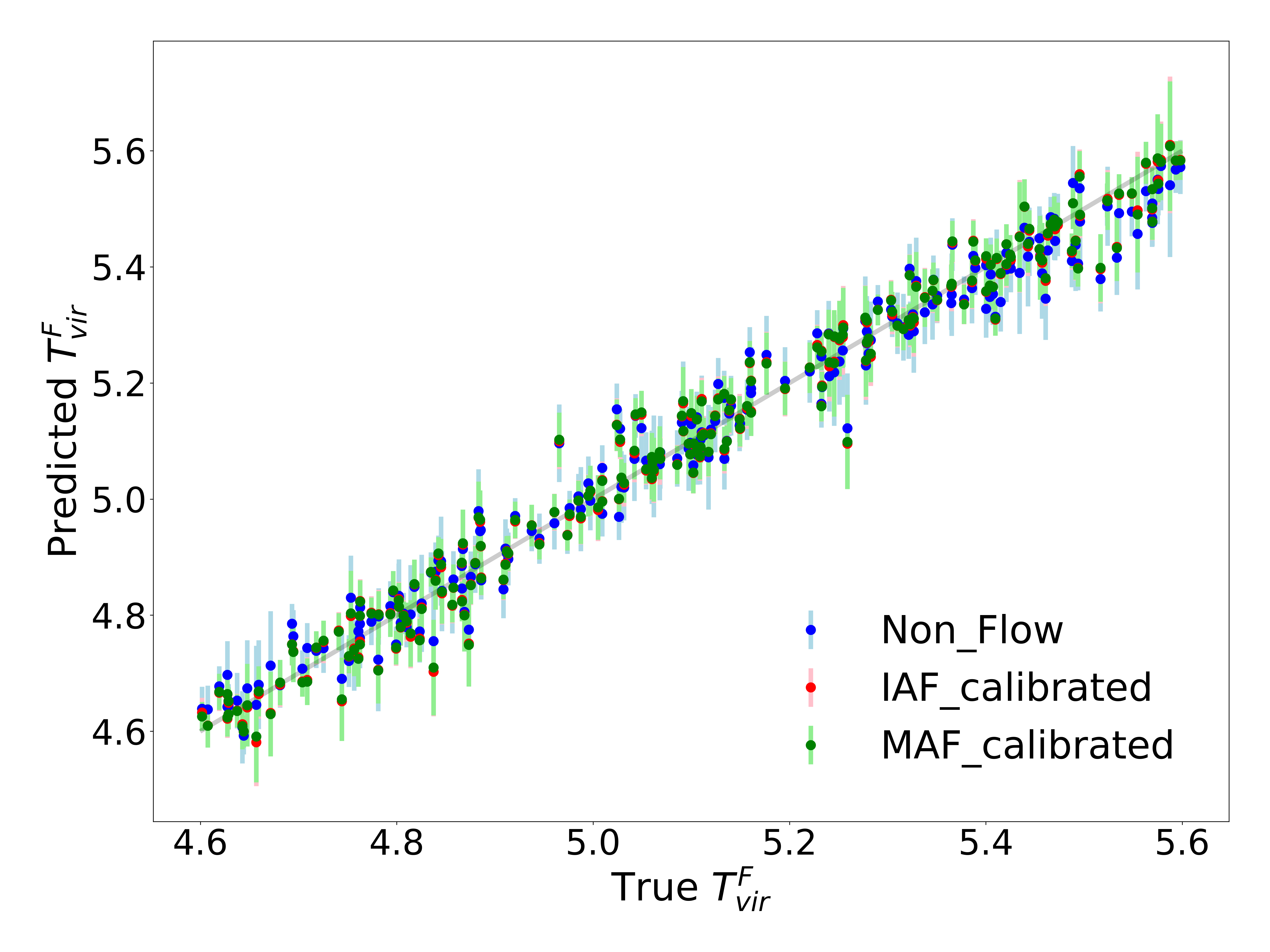}
        \caption{Predicted $T^F_{vir}$ value vs true value.}
        \label{figpredd}
    \end{subfigure}
    \caption{Parameter inference for different Flow methods using the test set after calibration. The images in each panel contain the true values vs the predicted parameter values. The shadow lines represent the total uncertainty, epistemic plus aleatoric.}\label{figapendix1flow}
\end{figure}

In order to compare the methods used so far, we chose an example from the test set, and produced the posterior and marginalized distributions of the parameters, obtaining the results displayed in Fig.~\ref{fig:bestCIspflowcal}. We can observe that after calibration, the contours produced by MAF becomes wider solving the underestimation found during training. Moreover, the contours of MAF and IAF applied in calibration overlap, and they are smaller compared with the base experiment, while Flows applied during training produce better results only for IAF. The credible intervals at  $95\%$ are shown in Table~\ref{table:bestcalflowcal}. There we can see the effect of the flows on the performance of the network, allowing for a reduction of the uncertainty intervals.

Finally, in Fig.~\ref{figapendix1flow} we plot the predicted and true values of the cosmological and astrophysical parameters using the model calibrated with IAF and MAF Flows. The error bars displayed in the plots correspond to both aleatoric and epistemic uncertainties.
Here we observe that $\sigma_8$ parameter contains larger errors  which means this parameter is the most difficult
to predict accurately, this could be a consequence of the 21cm signal being less sensitive to the effects of the density field than the IGM properties~\cite{10.1093/mnras/stz010,hassan2019constraining}.
The ionizing efficiency, $\zeta$, presents instead accurate predictions at low values, which are getting progressively less accurate and less precise at larger values.
This fact could be explained due to the limited information at high  redshifts (which can be reduced by assuming not spin temperature saturation) and also,  the small variability of the brightness temperature maps at lower redshift with respect to large values of $\zeta$.

\section{Conclusion and discussion}\label{sec:conclus}
We presented the first study using  Bayesian Neural Network and Normalizing Flows to obtain credible estimates for astrophysical and cosmological parameters from 21cm signals. These  methods offer alternative ways different from  MCMC to make inference and recover the information in the 21 cm observations.
Firstly, we show that Flipout outperforms Dropout
and is able both to better estimate parameters correlations and to obtain a better coefficient of determination.
Comparing with existing literature, we obtain comparable performances, while using a relatively smaller network~\cite{10.1093/mnras/stz010,hassan2019constraining}, furthermore by using a BNN, based on Variational Inference, we can estimate the confidence intervals for the predictions and the parameters uncertainties correlations.
The 21 cm signal is highly non-Gaussian due to the complex physics involved during the EoR. Normalising Flows provide a flexible likelihood model capable to better capture complex information encoded in the dataset. This improves the performance of the network, training with flows achieves better NLL values than experiments without Flows (Tables~\ref{table:bestCI} and ~\ref{table:bestCIflow}).
Additionally we propose novel calibration methods employing flows after training, showing how this method provide accurate uncertainty estimates and high prediction of the parameters regardless of the flow used (Table~\ref{table:bestcalflowcal}).
Fine tuning the last layer, in combination with NFs leads to a  simple, fast and  effective technique for calibrating BNNs.

As future perspective, we plan on evaluating the performances of different  network architectures (also in particular residual networks) and estimate the cosmological and astrophysical parameters in the presence of realistic noise from instruments of the future 21 cm surveys and also in other astrophysical dataset.

\section*{Acknowledgments}
H.J.~Hort\'ua, R.~Volpi, and L.~Malag\`o are supported by the DeepRiemann project, co-funded by the European Regional Development Fund and the Romanian Government through the Competitiveness Operational Programme 2014-2020, Action 1.1.4, project ID P\_37\_714, contract no. 136/27.09.2016. 

%References are before Appendix
%\section*{References}
\bibliographystyle{unsrt}
\bibliography{paper21cm}

%Appendix is after everything
\appendix
\section{Appendix}
In Tables~\ref{table:Dropoutall}-\ref{table:Flipoutall} we report different experiments, to determine the most adequate technique for estimating the parameters from the 21cm dataset. First, we found that sampling more than once during training improves the results. Second, Flipout does a good job for extracting the information in the 21cm images rather than other techniques such as Dropout.

\begin{figure}[h!]
\begin{center}
%\framebox[4.0in]{$\;$}
\includegraphics[width=0.8\linewidth]{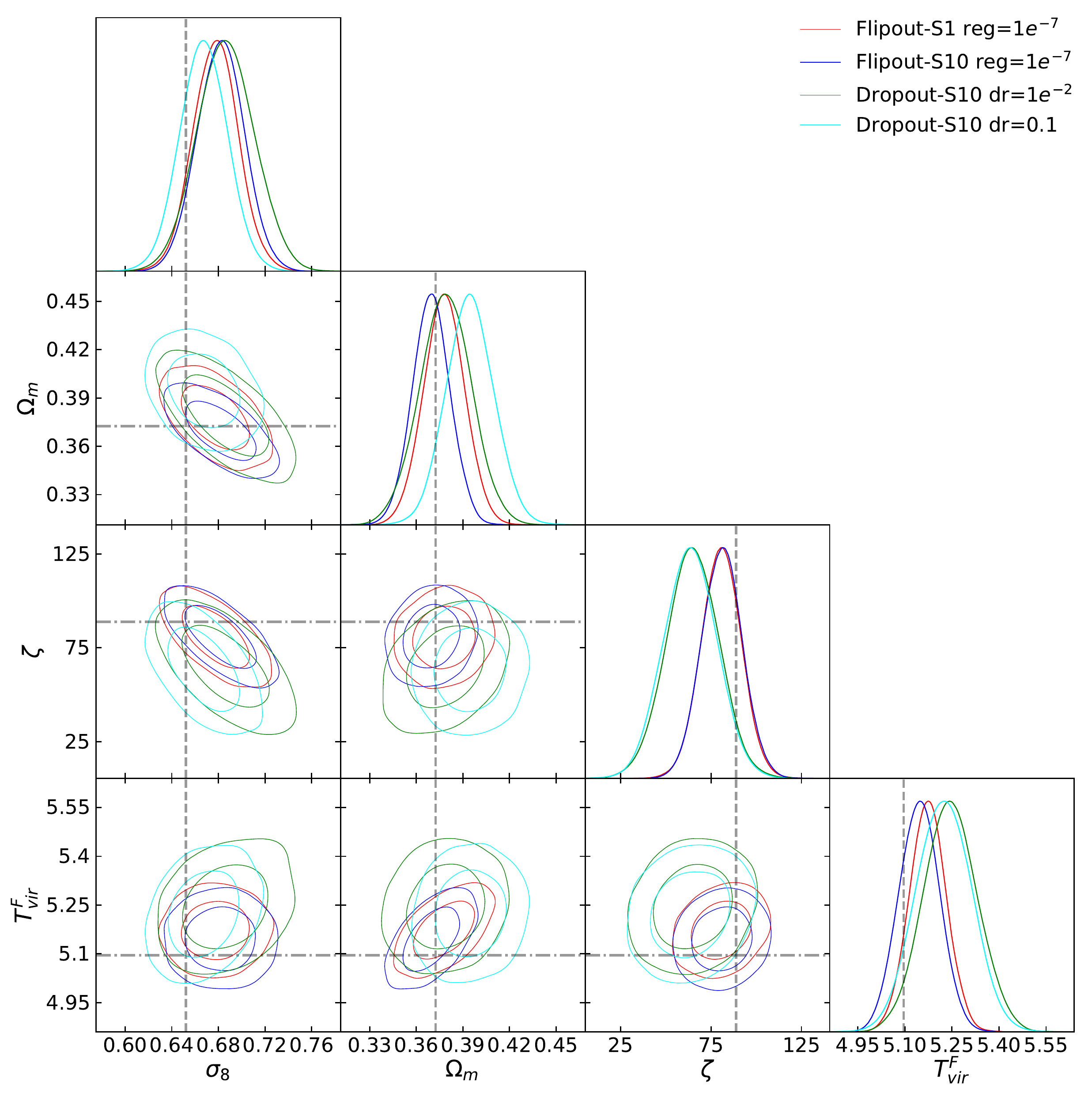}
%\fbox{\rule[-.5cm]{0cm}{4cm}\rule[-.5cm]{4cm}{0cm}}
\end{center}
\caption{One and two-dimensional posterior distributions of the parameters for one example in the test set. Each color represents different experiments.}
\label{fig:all4CI}
\end{figure}

 \begin{table}[h!]
\centering
\begin{tabular}{|l|l|l|l|l|}
\hline
 & \multicolumn{2}{l|}{Dropout dr$=1e^{-2}$} & \multicolumn{2}{l|}{Dropout dr$=0.1$} \\ \cline{2-5} 
                  & Sample=1          & Sample=10          & Sample=1          & Sample=10          \\ \hline
NLL               &    -0.18        & -0.74            &  0.99          & 0.28            \\ \hline
$R^2$                &     0.77       & 0.85            &      0.70      &    0.78         \\ \hline
$68\%$ C.L.         &    65.3        &    68.1         &       66.7     &     65.0        \\ \hline
$95\%$ C.L.     &    94.1        &       95.5      &     92.8       &     92.2        \\ \hline
$99\%$ C.L.     &       99.3     &      99.7       &       98.7     &     98.7        \\ \hline
\end{tabular}
\caption{Metrics for all Dropout experiments: dr$=(1e^{-2},0.1)$, reg $=1e^{-5}$. In each experiment, we sample once and ten times during training.}
\label{table:Dropoutall}
\end{table}

\begin{table}[h!]
\centering
\begin{tabular}{|l|l|l|l|l|}
\hline
 & \multicolumn{2}{l|}{Flipout reg$=1e{-7}$} & \multicolumn{2}{l|}{Flipout reg$=1e^{-5}$} \\ \cline{2-5} 
 & Sample=1           & Sample=10          & Sample=1           & Sample=10 \\ \hline
NLL               &    -2.30         & -2.94            &     -1.81        &-2.00   \\ \hline
$R^2$                &    0.91         &   0.94          &     0.84        & 0.84   \\ \hline
$68\%$ C.L.                &     75.5        &  73.0           &     76.2        &  76.4    \\ \hline
$95\%$ C.L.                &       97.2      &  96.8           &     97.6        &      97.5    \\ \hline
$99\%$ C.L.                &      99.5       &   99.8     &    99.8         &      99.8    \\ \hline
\end{tabular}
\caption{Metrics for all Flipout experiments: reg $=(1e^{-5}, 1e^{-7})$. In each experiment, we sample once and ten times during training.}
\label{table:Flipoutall}
\end{table}
Finally, we observe that Dropout underestimates its uncertainties while Flipout overestimates its uncertainties, therefore methods for calibration should be used before reporting the predictions. The confidence level reported in Tables~\ref{table:Dropoutall}-\ref{table:Flipoutall}, are computed with the method explained in Sec.~\ref{calibrat}. The contour regions for the best results are also reported in Fig.~\ref{fig:all4CI}. The $R^2$ for Flipout is reported  in Table~\ref{table:Flipoutall}.

\end{document}